%% file: ms.tex
\pdfoutput=1
\documentclass{aastex62}

\usepackage{ifthen}
\usepackage{etoolbox}

\usepackage{longtable}
\usepackage{graphicx}
\usepackage{amsmath,amssymb}
\usepackage{color}
\usepackage{units}
\usepackage{epstopdf}
\usepackage{hyperref}
\usepackage{url}

\newcommand{\beq}{\begin{equation}}
\newcommand{\eeq}{\end{equation}}
\newcommand{\bdm}{\begin{displaymath}}
\newcommand{\edm}{\end{displaymath}}

\definecolor{Gray}{gray}{0.9}

\graphicspath{{./}}

\newtoggle{endauthorlist}
\toggletrue{endauthorlist}

\begin{document}

\title{Search for transient gravitational wave signals associated with magnetar bursts during Advanced LIGO's second observing run}

\iftoggle{endauthorlist}{
  %
  %
  \let\mymaketitle\maketitle
  \let\myauthor\author
  \let\myaffiliation\affiliation
  \author{The LIGO Scientific Collaboration}
  \author{The Virgo Collaboration}
}{
  %
  %
}

\begin{abstract}
  We present the results of a search for short and intermediate-duration gravitational-wave signals from four magnetar bursts in Advanced LIGO's second observing run. We find no evidence of a signal and set upper bounds on the root sum squared of the total dimensionless strain ($h_{\text{rss}}$) from incoming intermediate-duration gravitational waves ranging from $1.1 \times 10^{-22}$ at 150\,Hz to $4.4 \times 10^{-22}$ at 1550\,Hz at 50\% detection efficiency. From the known distance to the magnetar SGR 1806-20 (8.7 kpc) we can place upper bounds on the isotropic gravitational wave energy of $3.4 \times 10^{44} \text{erg}$ at 150\,Hz assuming optimal orientation. This represents an improvement of about a factor of 10 in strain sensitivity from the previous search for such signals, conducted during initial LIGO's sixth science run. The short duration search yielded upper limits of $2.1 \times 10^{44} \,\text{erg}$ for short white noise bursts, and $2.3\times 10^{47}\,\text{erg}$ for $100 \, \text{ms}$ long ringdowns at 1500\,Hz, both at 50\% detection efficiency.
\end{abstract}


\section{Introduction}

So far, the Laser Interferometer Gravitational-Wave Observatory (LIGO) \citep{aligo} and Virgo \citep{avirgo}, have reported detections of a handful of gravitational-wave (GW) signals from coalescence of compact binary systems \citep{AbEA2016a,AbEA2016e,AbEA2017a,AbEA2017b,AbEA2017c,AbEA2017d}. Isolated compact objects may also emit detectable GWs, though they are predicted to be much weaker than compact binary coalescences \citep{Sathyaprakash2009}. Because of the high energies and mass densities required to generate detectable GWs, neutron stars and supernovae are among the main targets of non-binary searches. This paper focuses on a type of neutron star: magnetars.

Magnetars are highly magnetized isolated neutron stars \citep{Mereghetti2015,woods_thompson_2006}. Originally classified as Anomalous X-ray Pulsars (AXPs), or Soft Gamma Repeaters (SGRs), some AXPs were observed acting like SGRs and vice versa. They are now considered to be a single class of objects defined by their power source: the star's magnetic field, which, at $10^{13}-10^{15}\,\text{G}$, is about 100$\times$ stronger than a typical neutron star. Magnetars occasionally emit short bursts of soft $\gamma$-rays, but the exact mechanism responsible for the bursts is unclear. There are currently 23 known magnetars (and an additional 6 candidates) \citep{mcgillpaper},\footnote{See the catalog at \url{http://www.physics.mcgill.ca/\~pulsar/magnetar/main.html}} which were identified based on observations across wavelengths of bursts, continuous pulsating emission, spin-down rates, and glitches in their rotational frequency. The bursts last $\sim 0.1 \text{s}$ with luminosity of up to $\sim 2 \times 10^{42} \, \text{erg} \, \text{s}^{-1}$, and can usually be localized well enough to allow identification of the source magnetar. Many magnetars also emit pulsed X-rays and some are visible in radio. 

The large energies involved originally led to the belief that magnetar bursts could be promising sources of detectable gravitational waves, e.g. \cite{ioka01, corsi11}. Further theoretical investigation indicates that most mechanisms are likely too weak to be detectable by current detectors \citep{LevHov2011, ZinLas2012}. Nevertheless, due to the large amount of energy stored in their magnetic fields and known transient activity, magnetars remain a promising source of GW detections for ground-based detectors with rich underlying physics.

The search presented in this paper is triggered following identification of magnetar bursts by $\gamma$-ray telescopes. The methodology is similar to one done during initial LIGO's sixth science run \citep{QuBr2017, RQJthesis} with a few improvements and the use of an additional pipeline targeted toward shorter-duration signals (X-Pipeline) \citep{SuJo2010}. This pipeline has been used to look for GWs coincident with $\gamma$-ray bursts (GRBs) (see \cite{xgrbo1} for such searches during advanced LIGO's first observing run). 

The first searches for GW counterparts from magnetar activity targeted the 2004 hyperflare of SGR 1806-20. Initial LIGO data was used to constrain the GW emission associated with the quasi-periodic seismic oscillations(QPOs) of the magnetar following this catastrophic cosmic event \citep{Matone_2007, PhysRevD.76.062003} as well as the instantaneous gravitational emission \citep{Kalmus_2007, 2008_sgr_search}. \cite{2008_sgr_search} and \cite{AbAb2011} report on GW emission limits associated with additional magnetar activity observed during the initial detector era until June 2009. LIGO data coinciding with the 2006 SGR 1900+14 storm was additionally analyzed by “stacking” GW data \cite{kalmus_2009} corresponding to individual bursts in the storm’s EM light curve \citep{stacked_flares}. Additionally, a magnetar was considered as a possible source for a GRB during initial LIGO (GRB 051103), and a search using X-Pipeline and the Flare pipeline placed upper limits on GW emission from the star's fundamental ringing mode \citep{GRB051103}.

The rest of this paper is laid out as follows: in Section \ref{magbursts}, we provide a brief overview of the astrophysics of magnetars as is relevant to gravitational-wave astronomy and the short bursts used in this analysis. Next, in Section \ref{method} we describe the methodology of the GW search. Section \ref{results} describes the results and upper limits on possible gravitational radiation from the studied bursts. The Appendix contains a discussion of the effect of GW polarization on the sensitivity of the intermediate duration search.

\section{Magnetar Bursts} \label{magbursts}

Magnetars are currently not well understood. Their magnetic fields are strong and complex \citep{braithwaite}, and power the star's activity. Occasionally and unpredictably, magnetars give off short bursts of $\gamma$-rays whose exact mechanism is unknown, but may be caused by seismic events, Alfv\'en waves in the star's atmosphere, magnetic reconnection events, or some combination of these, e.g. \cite{thompson1995}. After some of the brighter bursts (giant flares, which have been seen only three times), there is a soft X-ray tail which lasts for hundreds of seconds. Quasi-periodic oscillations (QPOs) have been observed in the tail of giant flares \citep{IsBe2005,StrWat2005} and some short bursts \citep{short_burst_qpo_a, short_burst_qpo_b}, during which various frequencies appear, stay for hundreds of seconds, and then disappear again, indicating a resonance within the magnetar. Many possible resonant modes in the core and crust of the magnetar have been suggested to cause the QPOs, although it is unclear which modes actually produce them. Some of these modes, such as f-modes and r-modes, couple well to GWs, and so, if sufficiently excited, could produce detectable GWs, though current models indicate that they will be too weak \citep{LevHov2011,ZinLas2012}. Other modes, such as the lowest order torsional mode, do not create the time changing quadrupole moment needed for GW emission. None of these models provide precise predictions for emitted GW waveforms.

This search was performed on data coincident with the four short bursts from magnetars during advanced LIGO's second observing run for which there was sufficient data (we require data from two detectors) for both short (less than a second long) and intermediate (hundreds of seconds long) duration signals. Table \ref{burst_data} describes the four bursts. In addition to the four studied bursts, there were five bursts that occurred during times when at least one detector was offline. No GW analysis was done on them. All GW detector data comes from the two LIGO detectors because Virgo was not taking data during any of these bursts.

Three bursts come from the magnetar SGR 1806-20. They were all identified by the Burst Alert Telescope (BAT) aboard NASA's Swift satellite \citep{swift2004}. These were sub-threshold events that were found in BAT data (D. M. Palmer, personal communication, June 6, 2017), an example of which is shown in Figure \ref{SWIFT_data}, with the data from the other two found in Appendix \ref{swiftplots}. The fourth was a short GRB with a soft spectrum observed by the \textit{Fermi} Gamma-ray Burst Monitor \citep{fermi2009}, and named GRB 170304A.

\begin{table}[h!]
\centering
 \begin{tabular}{|c c c c c c||} 

 \hline
 Source & Date  & Time & Duration & Fluence & Distance \\ 
 & & (UTC) & (s) & (erg $\text{cm}^{-2}$) & (kpc) \\[0.5ex]
 \hline\hline
 SGR 1806-20 & Feb 11, 2017 & 21:51:58 & 0.256 & $8.9 \times 10^{-11}$ & 8.7 \\ 
 \hline
 SGR 1806-20 & Feb 25, 2017 & 06:15:07 & 0.016 & $1.2 \times 10^{-11}$ & 8.7 \\
  \hline
 GRB170304A & March 4, 2017 & 00:04:26 & 0.16 & $3.1 \times 10^{-10}$ & --\\
  \hline
 SGR 1806-20 & April 29, 2017 & 17:00:44 & 0.008 & $1.4 \times 10^{-11}$ & 8.7\\[1ex] 
 \hline

\end{tabular}
\caption{List of magnetar bursts considered in this GW search. GRB170304A is described in GCN circular 20813; data on SGR 1806-20 burst activity is courtesy of David M. Palmer.}
\label{burst_data}
\end{table}

\begin{figure}
\centering
\includegraphics[width=0.5\textwidth, angle=90]{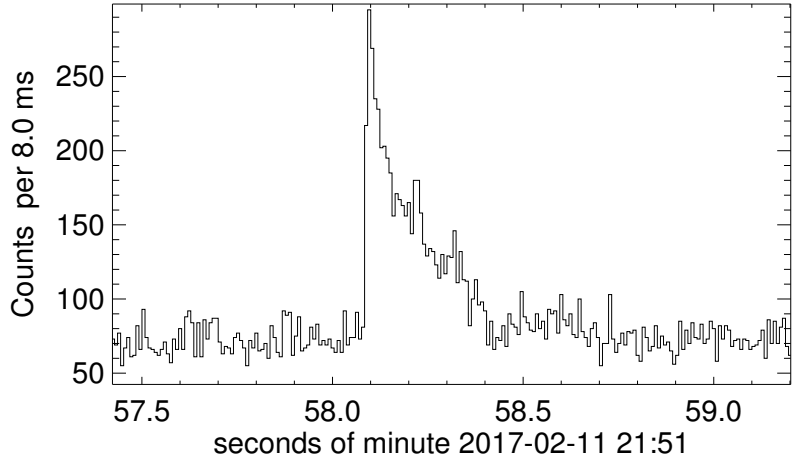}
\caption{\label{SWIFT_data}SWIFT BAT's data for the February 11 burst from SGR 1806-20. Image courtesy of David M. Palmer.}
\end{figure}

\section{Method} \label{method}

\subsection{Excess power searches}

Fundamentally, all multi-detector GW searches seek to identify GW signals that are consistent with the data collected at both detectors. Some searches identify candidate signals in each detector separately, then later consider only the candidates that occur in all detectors within the light-travel time and with the same signal parameters. This approach is disfavored in searches that do not rely on templates. We cannot perform a templated search here because there is no current model which can produce templates for magnetar GW bursts. Instead, we first combine the two data streams to create a time-frequency map where the value in each time-frequency pixel represents some measure of the GWs (often energy) consistent with the observations from the detectors.

The next step is to identify GW signals in the time-frequency map. This is done by clustering together groups of pixels, calculating the significance of each cluster with a metric, and searching for the most significant cluster. In order to cover a broader range of frequencies and time scales, we use two different analysis pipelines which use different clustering algorithms. 

The short-duration search uses seed-based clustering implemented by X-Pipeline, which focuses on groups of bright pixels (the seed) \citep{SuJo2010}. Specifically, the clusters considered by X-Pipeline are groups of neighboring pixels that are all louder than a chosen threshold. This approach works well for short-duration searches, but fails for longer-duration signals for two reasons: random noise will tend to break up the signal into multiple clusters, and each pixel is closer to the background, so fewer of them will be above the threshold.

We rely on STAMP \citep{ThKa2011} for the intermediate-duration search. STAMP offers a seedless method whose clustering algorithm integrates over many, randomly chosen, B\'ezier curves \citep{ThCo2013, ThCo2014}. Because of this, it can jump over gaps in clusters caused by noise, and thus it is better suited for longer-duration signals. Additionally, it can build up signal-to-noise ratio (SNR) over many pixels of only slightly elevated SNR. This method was previously used to search for signals from magnetars during initial LIGO \citep{QuBr2017, RQJthesis}. 

\subsection{X-Pipeline}

X-Pipeline is a software package designed to search for short-duration gravitational wave signals in multiple detectors, and includes automatic glitch rejection, background calculation, and software injection processing (for details, see \cite{SuJo2010}). It forms coherent combinations from multiple detectors, thus making it relatively insensitive to non-GW signals, such as instrumental artifacts. X-Pipeline is used primarily to search for GWs coincident with $\gamma$-ray bursts (GRBs), but is suitable for any short-duration coherent search.

X-Pipeline takes a likelihood approach to estimating the GW energy found in each time-frequency pixel. It models the data collected at the detectors as a combination of signal and detector noise, then uses a maximum likelihood technique to calculate the estimated GW signal power in each time-frequency pixel.

For clustering, X-Pipeline selects the loudest 1\% of pixels and connects neighboring pixels. Each connected group is a cluster, and the clusters are scored based on the likelihood described in the previous paragraph. We want to pick the time length of the pixels in the time-frequency map so that signal is present in the smallest number of time-frequency pixels, as this will recover the signal with the highest likelihood. Since we do not have a model for the waveform we are searching for, we use multiple pixel lengths and run the clustering algorithm on all of them. After clusters are identified, X-Pipeline identifies which candidate clusters are likely glitches by comparing three measurements of signal energy: coherent energy consistent with GWs, coherent energy inconsistent with GWs, and sum of the signal energy in all detectors (referred to as incoherent energy). GW signals can be differentiated from noise by the ratio of coherent energy inconsistent with GWs to the incoherent energy (see Sections 2.6 and 3.4 of \cite{SuJo2010} for full details).

The primary target of this search are GWs produced from the excitation of the magnetar's fundamental mode, which are primarily dampened by the emission of GWs \citep{Det1975,AndKok1998}. We have chosen parameters for X-Pipeline to search for signals a few hundred milliseconds long. The search window begins 4 seconds before the $\gamma$-rays arrive and ends 4 seconds after. The frequency range for the short-duration search is 64--4000\,Hz, and the pixel lengths are every factor of 2 between 2\,s and 1/128\,s, inclusive.

\subsection{STAMP}

STAMP is an unmodeled, coherent, directed excess power search suitable for longer-duration signals, described in more detail in \cite{ThKa2011}. In short, the pipeline calculates the cross power between the two detectors, accounting for the time delay due to the light travel time between detectors. It then makes this into a time-frequency SNR map, where pixel SNR is estimated from the variable $\big \langle \hat{Y} (t;f,\hat{\Omega}, )\big \rangle = \text{Re} \Big[ \tilde{Q}_{IJ}(t;f,\hat{\Omega}) \big(2 \, \tilde{s}_I^*(t;f) \tilde{s}_J(t;f) \big) \Big]$. Here, $\tilde{s}_I(t;f)$ is the Fourier transformed data from detector $I$ and $\tilde{Q}_{IJ}(t; f, \hat{\Omega})$ is the filter function required given the locations and orientations of the pair of detectors and the sky position $\hat{\Omega}$ of the source \citep{ThKa2011}. The ideal filter function also depends on the polarization of the incoming GWs, which is unknown. Thus the best we can do is to use the unpolarized filter function which causes a loss of signal power (though this loss is nearly zero for optimal sky locations). This is more fully discussed in Appendix \ref{polarization}.

To identify signals, STAMP uses seedless clustering and searches over a large number of clusters (30 million in this search, see section III of \citep{ThCo2013}). For clusters, we use B\'ezier curves, which are parameterized by three points \citep{ThCo2013}.
 
GWs radiated through the mechanisms related to QPOs would be monochromatic, or close to it. So, it makes sense to only search for such signals. Through STAMP, this is easily accomplished by restricting the search to clusters whose frequencies change by only a small amount. This reduces the number of possible clusters, which means we get all of the benefits of searching more clusters without the additional computational cost. Restricting the frequency change too much may cause signals to be missed completely. Compromising between these, we restrict the searched clusters to those with a frequency change less than 10\%.

\begin{figure}[ht!]
\centering
\includegraphics[width=0.7\textwidth]{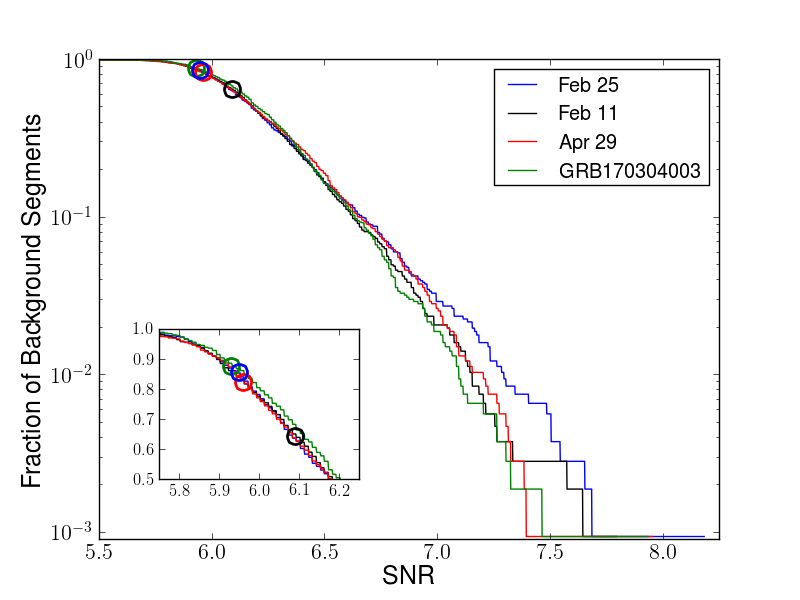}
\caption{\label{background}SNR distribution of the background (lines) and on-source result (open circles) for each burst for the intermediate-duration search. As expected, the background distributions are similar; since many background analyses give louder SNR than the on-source, we conclude that no signal has been detected. Inset: a detailed view of the on-source results.}
\end{figure}

To estimate the background for this search, we use approximately 15 hours of data from each detector collected around the time of each burst, excluding the data, coincident with the burst, that was searched for GWs (the on-source). We can then perform background experiments free of any possible coincident GW signal by pairing data taken at different times as if they were coincident. Then, breaking up the background data into 33 segments, we generate 1,056 background experiments. Each of these is run in exactly the same way as the on-source analysis, giving an estimate of the SNR that can be expected from detector noise. The resulting distributions for the background data for each burst, along with the on-source results, are shown in Figure \ref{background}.

With unlimited computing power, we would calculate upper limits by adding software injections of increasing amplitude until the desired fraction of injections are recovered. However, this is prohibitively expensive in computing time.
Instead, at each amplitude for the injection, we search only over a few previously identified clusters. To pick those clusters we do a full run using the seedless algorithm with fifteen injections at varied amplitudes around the expected recovery threshold and identify clusters which recover the injection, setting the same random seed as was used for on-source recovery.  We then analyze a large number of injections using only the pre-identified clusters, and calculate the maximum SNR of those clusters. The injections at all amplitudes are done with the same time-frequency parameters. This allows us to efficiently recover the injection by searching a small fraction of the total clusters. This is shown to work as expected in Appendix \ref{singletrack}.

Since choosing a random seed also chooses which clusters will be searched, this value can affect the final upper limit values. This effect is limited by using a large number of clusters (30 million), and analysis performed with different seeds shows this effect leads to about a 10\% uncertainty in the resulting upper limit value. However, this is not a completely new source of error - changing the seed is functionally equivalent to changing the time-frequency location of the software injections. In addition, the on-source analysis uses only one seed since it is only run once, so this uncertainty is a manifestation of the random nature of the search.

\section{Results and Discussion} \label{results}

No signals were found by either the short- or intermediate-duration searches. We present the results and upper limits on GW strain and energy for each analysis below.

\subsection{Short-Duration Search Upper Limits}

No significant signal was found by X-Pipeline. After glitch rejection, the most significant cluster for the February 25 burst had a p-value of 0.63.

Following the previous f-mode search \citep{GRB051103}, we injected white noise bursts (frequencies: 100--200\,Hz and 100--1000\,Hz; durations: 11\,ms and 100\,ms), and ringdowns (damped sinusoids, at frequencies: 1500\,Hz and 2500\,Hz; time constants: 100\,ms and 200\,ms), and chirplets (chirping sine-Gaussians; this differs from the prior search, which used sine-Gaussians). The best limits for the white noise bursts were for the 11 ms long bursts in the 100--200 Hz band, at $2.1 \times 10^{44} \text{ erg}$ in total isotropic energy and $h_{\text{rss}}$ of $5.6 \times 10^{-23}$ at the detectors. We are most sensitive to ringdowns at 1500 Hz and a time constant of 100 ms, with an upper limit of $2.3 \times 10^{47} \text{ erg}$ and $h_{\text{rss}}$ of $1.9 \times 10^{-22}$. Directly comparing the $h_{\text{rss}}$ limits to \cite{AbAb2011}, we see that limits have improved by roughly a factor of 10, though the ringdowns we used had slightly different parameters. Comparing to \cite{GRB051103}, which provided only energy upper limits assuming a distance of 3.6 Mpc, we see an improvement of factor of 60 after correcting for the larger distance. This corresponds to roughly a factor of 8 improvement in $h_{\text{rss}}$ limits. A full list of upper limits for the waveforms tested is found in Table \ref{x_lims}.

\begin{table}[ht!]
\centering
 \begin{tabular}{||c c c c c||} 

 \hline
 Injection Type & Frequency (Hz) & Duration/ & $h_{\text{rss}}$ & Energy (erg) \\ 
 & & $\tau$ (ms)& & \\[0.5ex] 
 \hline\hline
 chirplet & 100 & 10 & $5.42\times 10^{-23}$ & $8.49\times 10^{43}$ \\
\hline
chirplet & 150 & 6.667 & $4.93\times 10^{-23}$ & $1.58\times 10^{44}$ \\
\hline
chirplet & 300 & 3.333 & $5.29\times 10^{-23}$ & $7.27\times 10^{44}$ \\
\hline
chirplet & 1000 & 1 & $1.15\times 10^{-22}$ & $3.82\times 10^{46}$ \\
\hline
chirplet & 1500 & 0.6667 & $1.69\times 10^{-22}$ & $1.81\times 10^{47}$ \\
\hline
chirplet & 2000 & 0.5 & $2.32\times 10^{-22}$ & $5.92\times 10^{47}$ \\
\hline
chirplet & 2500 & 0.4 & $3.06\times 10^{-22}$ & $1.56\times 10^{48}$ \\
\hline
chirplet & 3000 & 0.3333 & $3.96\times 10^{-22}$ & $3.65\times 10^{48}$ \\
\hline
chirplet & 3500 & 0.2857 & $5.30\times 10^{-22}$ & $8.51\times 10^{48}$ \\
\hline
white noise burst & 100--200 & 11 & $5.57\times 10^{-23}$ & $2.09\times 10^{44}$ \\
\hline
white noise burst & 100--200 & 100 & $7.88\times 10^{-23}$ & $4.15\times 10^{44}$ \\
\hline
white noise burst & 100--1000 & 11 & $1.00\times 10^{-22}$ & $1.04\times 10^{46}$ \\
\hline
white noise burst & 100--1000 & 100 & $1.83\times 10^{-22}$ & $3.55\times 10^{46}$ \\
\hline
ringdown & 1500 & 200 & $1.89\times 10^{-22}$ & $2.25\times 10^{47}$ \\
\hline
ringdown & 2500 & 200 & $2.87\times 10^{-22}$ & $1.37\times 10^{48}$ \\
\hline
ringdown & 1500 & 100 & $1.89\times 10^{-22}$ & $2.25\times 10^{47}$ \\
\hline
ringdown & 2500 & 100 & $2.80\times 10^{-22}$ & $1.30\times 10^{48}$ \\

 \hline

\end{tabular}
\caption{Upper limits on isotropic energy from the short-duration search for the February 25 burst from SGR 1806-20. For white noise bursts, we give the duration of the injection; for the other waveforms, the characteristic time. All limits are given at 50\% detection efficiency, meaning that a signal with the given parameters would be detected 50\% of the time.}
\label{x_lims}
\end{table}

\subsection{Intermediate Duration Search Upper Limits}

To calculate upper limits, we add software injections of two waveforms (half-sine Gaussians and exponentially decaying sinusoids) at five frequencies (55, 150, 450, 750, and 1550 Hz) and at two timescales (150 seconds and 400 seconds). Reported upper limits are for 50\% recovery efficiency, where recovery is defined as finding a cluster, at the same time and frequency as the injection, with SNR greater than that of the on-source (for the February 25 event, it was 6.09). Full results are shown in Table \ref{stamp_lims}.

\begin{figure*}[ht!]
\centering
\includegraphics[width=\textwidth]{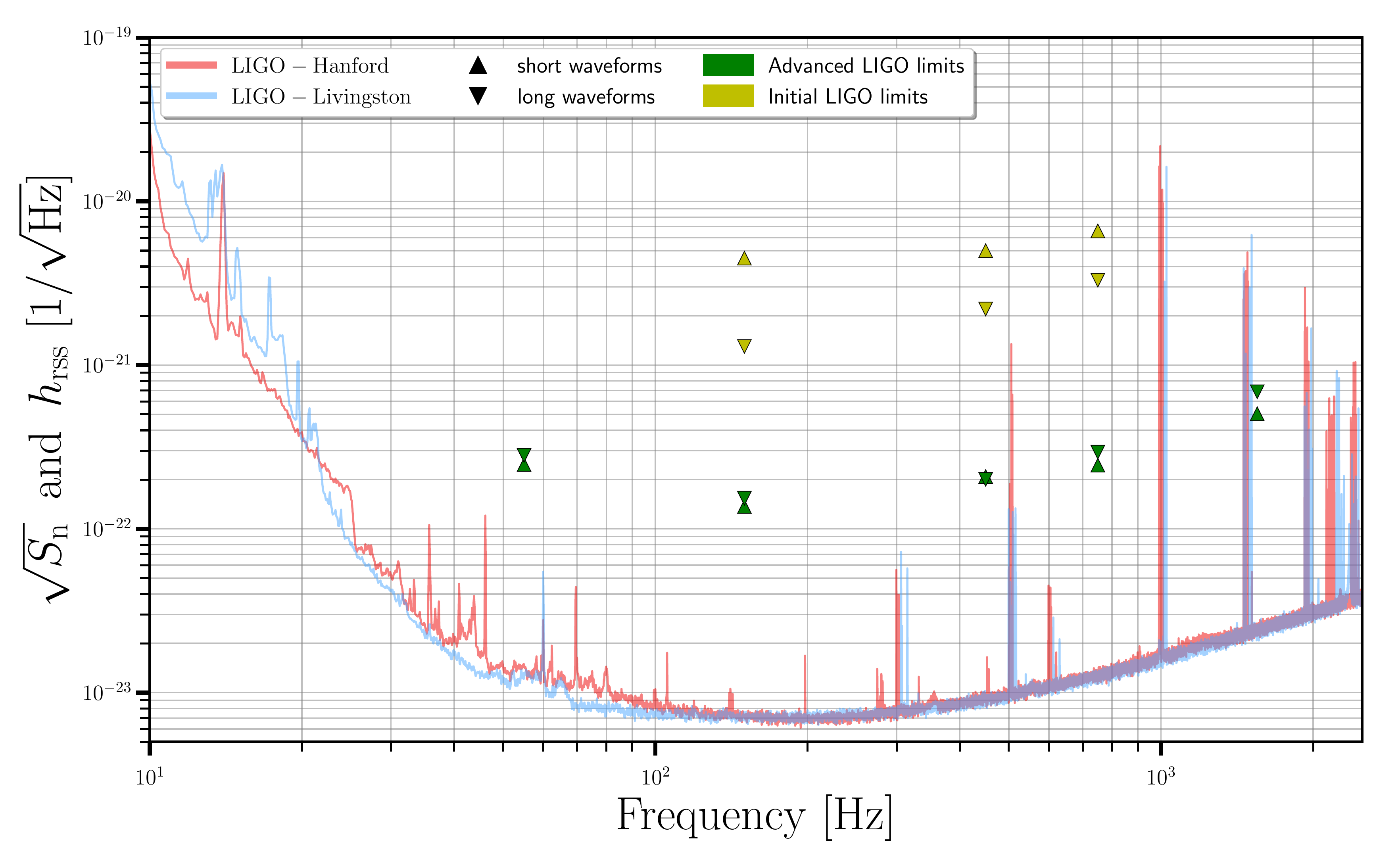}
\includegraphics[width=\textwidth]{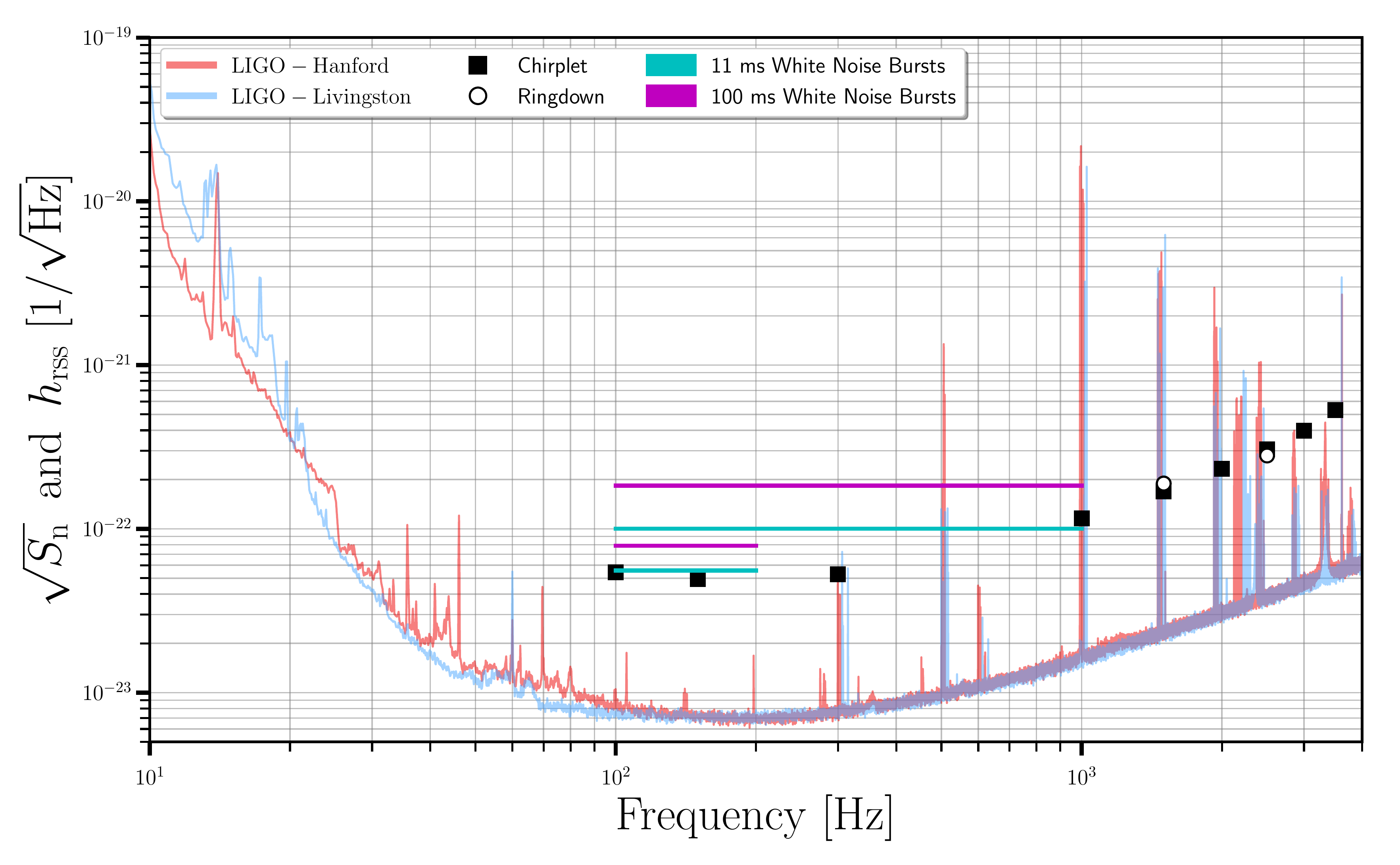}
\caption{\label{sensitivity}Upper limits for the the intermediate-duration search (above) and short-duration search (below), along with the sensitivity of the detectors. We plot $h_{\text{rss}}$ at 90\% detection efficiency for the intermediate-duration search here to allow direct comparison to published figures for the previous search in initial LIGO \citep{QuBr2017}. Short-duration limits are for 50\% efficiency as before. The advanced LIGO search limits are for the February 25 burst from SGR 1806-20 during the second observing run, and detector sensitivity is calculated from data during the analysis window.}
\end{figure*}

Due to the improved sensitivity of Advanced LIGO, we are able to set strain upper limits about a factor of 10 lower than the previous search during initial LIGO \citep{QuBr2017}, see Figure \ref{sensitivity}. Unlike the previous search, this search showed little difference in $h_{\text{rss}}$ sensitivity between the two injection lengths. STAMP has been refined to improve PSD estimation, which explains the small gap between the injection timescales for this search.

\begin{table}[ht!]
\centering
 \begin{tabular}{||c c c c c c||} 

 \hline
 Frequency (Hz) & Tau (sec) & \multicolumn{2}{c}{$h_{\text{rss}}$} & \multicolumn{2}{c||}{Energy (erg)}\\ 
  &  & Half Sine-Gaussian & Ringdown & Half Sine-Gaussian & Ringdown \\[0.5ex] 
 \hline\hline
55 & 400 & $2.29\times 10^{-22}$ & $2.43\times 10^{-22}$ & $1.82\times 10^{44}$ & $2.06\times 10^{44}$ \\
\hline
55 & 150 & $1.97\times 10^{-22}$ & $2.11\times 10^{-22}$ & $1.35\times 10^{44}$ & $1.55\times 10^{44}$ \\
\hline
150 & 400 & $1.32\times 10^{-22}$ & $1.37\times 10^{-22}$ & $4.52\times 10^{44}$ & $4.86\times 10^{44}$ \\
\hline
150 & 150 & $1.14\times 10^{-22}$ & $1.22\times 10^{-22}$ & $3.37\times 10^{44}$ & $3.89\times 10^{44}$ \\
\hline
450 & 400 & $1.69\times 10^{-22}$ & $1.79\times 10^{-22}$ & $6.62\times 10^{45}$ & $7.47\times 10^{45}$ \\
\hline
450 & 150 & $1.78\times 10^{-22}$ & $1.83\times 10^{-22}$ & $7.43\times 10^{45}$ & $7.83\times 10^{45}$ \\
\hline
750 & 400 & $2.56\times 10^{-22}$ & $2.70\times 10^{-22}$ & $4.21\times 10^{46}$ & $4.69\times 10^{46}$ \\
\hline
750 & 150 & $2.11\times 10^{-22}$ & $2.37\times 10^{-22}$ & $2.87\times 10^{46}$ & $3.61\times 10^{46}$ \\
\hline
1550 & 400 & $5.86\times 10^{-22}$ & $6.22\times 10^{-22}$ & $9.21\times 10^{47}$ & $1.03\times 10^{48}$ \\
\hline
1550 & 150 & $4.38\times 10^{-22}$ & $4.58\times 10^{-22}$ & $5.16\times 10^{47}$ & $5.62\times 10^{47}$ \\
 \hline

\end{tabular}
\caption{Upper limits on GW strain and energy from the intermediate-duration search for the February 25 burst from SGR 1806-20. All limits are at 50\% detection efficiency.}
\label{stamp_lims}
\end{table}

\subsection{Discussion}

This search has set the strongest upper limits on short- and intermediate-duration GW emission associated with magnetar bursts. The energy limits, which are as low as $10^{44}$--$10^{47} \text{erg}$, are now well below the EM energy scale of magnetar giant flares ($10^{46} \text{erg}$). The short bursts analyzed here were much weaker than a giant flare (see Table \ref{burst_data}), so for these bursts the limit is much larger than the observed electromagnetic energy. In addition, these limits assume ideal orientation of the magnetar (both sky position and polarization of produced GWs). The impact of other polarizations on the intermediate-duration search are discussed in Appendix \ref{polarization}, and plotted in Figure \ref{distance}.

\begin{figure}[ht!]
\centering
\includegraphics[width=\textwidth]{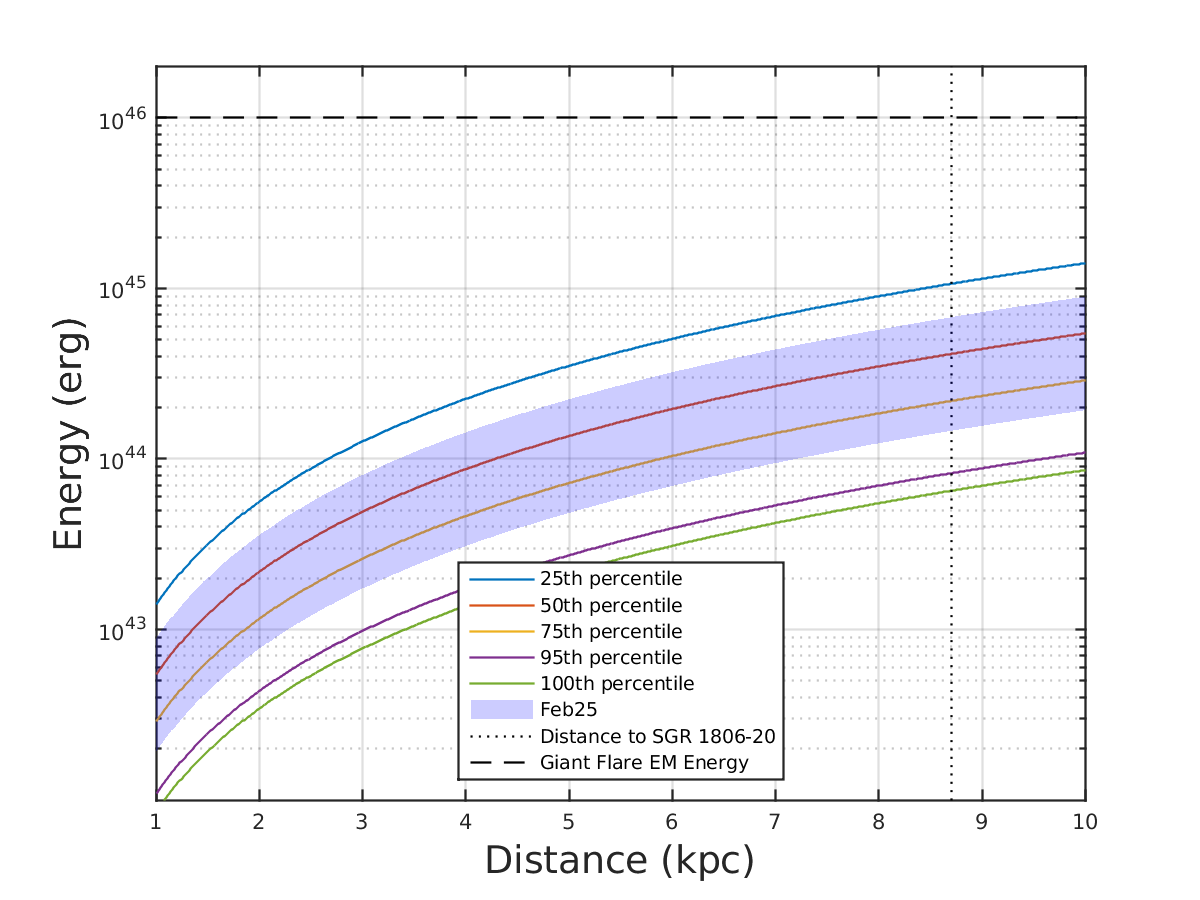}
\caption{\label{distance} Minimum detectable energy for the intermediate duration search vs distance for SGR 1806-20 for varied sky locations and GW polarizations at 55 Hz. The lines show how the variation in sky position (caused by the earth's rotation) and polarization (assumed to be random) affects the sensitivity; the purple 95\textsuperscript{th} percentile line indicates that the network sensitivity will be better than indicated by that line only 5\% of the time. The shaded region indicates the sensitivity to GW energy from the burst on February 25. Here, the uncertainty is only due to the unknown polarization.}
\end{figure}

The upper limits set by this search are still far above the GW energy from f-mode excitation during a giant flare according to \cite{ZinLas2012}, unless the magnetic field strength is far higher than currently accepted value of $2 \times 10^{15}\ \text{G}$ \citep{mcgillpaper}. Using Equation 2 from \cite{ZinLas2012}, f-mode GW emission from a giant flare would be about $1.4 \times 10^{38}\ \text{erg}$. A surface magnetic field of $1.8 \times 10^{16}\ \text{G}$ would be required to reach the best upper limit found with the short-duration source.

As the LIGO detectors increase in sensitivity, these upper limits will improve, and will be well-positioned to place meaningful limits on emitted GW energy in the event of a future nearby magnetar giant flare. 

\acknowledgements

\section*{Acknowledgements}
The authors gratefully acknowledge the support of the United States
National Science Foundation (NSF) for the construction and operation of the
LIGO Laboratory and Advanced LIGO as well as the Science and Technology Facilities Council (STFC) of the
United Kingdom, the Max-Planck-Society (MPS), and the State of
Niedersachsen/Germany for support of the construction of Advanced LIGO 
and construction and operation of the GEO600 detector. 
Additional support for Advanced LIGO was provided by the Australian Research Council.
The authors gratefully acknowledge the Italian Istituto Nazionale di Fisica Nucleare (INFN),  
the French Centre National de la Recherche Scientifique (CNRS) and
the Foundation for Fundamental Research on Matter supported by the Netherlands Organisation for Scientific Research, 
for the construction and operation of the Virgo detector
and the creation and support  of the EGO consortium. 
The authors also gratefully acknowledge research support from these agencies as well as by 
the Council of Scientific and Industrial Research of India, 
the Department of Science and Technology, India,
the Science \& Engineering Research Board (SERB), India,
the Ministry of Human Resource Development, India,
the Spanish  Agencia Estatal de Investigaci\'on,
the Vicepresid\`encia i Conselleria d'Innovaci\'o, Recerca i Turisme and the Conselleria d'Educaci\'o i Universitat del Govern de les Illes Balears,
the Conselleria d'Educaci\'o, Investigaci\'o, Cultura i Esport de la Generalitat Valenciana,
the National Science Centre of Poland,
the Swiss National Science Foundation (SNSF),
the Russian Foundation for Basic Research, 
the Russian Science Foundation,
the European Commission,
the European Regional Development Funds (ERDF),
the Royal Society, 
the Scottish Funding Council, 
the Scottish Universities Physics Alliance, 
the Hungarian Scientific Research Fund (OTKA),
the Lyon Institute of Origins (LIO),
the Paris \^{I}le-de-France Region, 
the National Research, Development and Innovation Office Hungary (NKFIH), 
the National Research Foundation of Korea,
Industry Canada and the Province of Ontario through the Ministry of Economic Development and Innovation, 
the Natural Science and Engineering Research Council Canada,
the Canadian Institute for Advanced Research,
the Brazilian Ministry of Science, Technology, Innovations, and Communications,
the International Center for Theoretical Physics South American Institute for Fundamental Research (ICTP-SAIFR), 
the Research Grants Council of Hong Kong,
the National Natural Science Foundation of China (NSFC),
the Leverhulme Trust, 
the Research Corporation, 
the Ministry of Science and Technology (MOST), Taiwan
and
the Kavli Foundation.
The authors gratefully acknowledge the support of the NSF, STFC, MPS, INFN, CNRS and the
State of Niedersachsen/Germany for provision of computational resources.

\bibliography{references}

\appendix

\section{Validation of the injection recovery method} \label{singletrack}

\begin{figure}[ht!]
\centering
\includegraphics[width=0.7\textwidth]{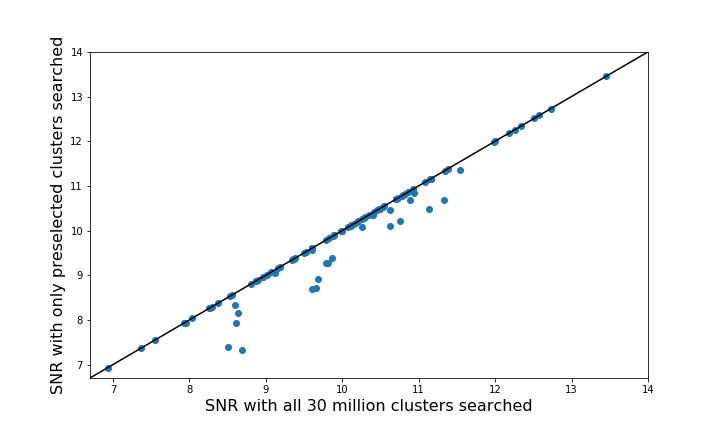}
\caption{\label{singletrack_comp} Direct comparison the SNR recovered by a full run of 30 million clusters and a run with only a few previously selected clusters. This shows that this method is highly effective in recovering the injected signal, at a small fraction of the computational cost.}
\end{figure}

In Section \ref{method}, we described the way we obtain the precise detectability threshold of injected waveforms. Given that this method is an approximation of the full search protocol, we check to see how closely the results match that of the full search. Figure \ref{singletrack_comp} shows the result of this test, run over 100 injection trials. A point falling below the diagonal indicates that the noise and injected signal conspire together in such a way that the cluster with the highest SNR is one that had not been previously identified. This test indicates that this happens infrequently. On average, we recover 98\% of the SNR that is found by the full search.

\section{STAMP Polarization} \label{polarization}

The cross power of a pixel in the time-frequency map is calculated with $\hat{Y}\equiv \text{Re} \Big[\widetilde{Q}_{IJ}(t;f,\hat{\Omega})C_{IJ}(t;f)\Big]$, where $\widetilde{Q}_{IJ}(t;f,\hat{\Omega})$ is the filter function which depends on the polarization of the GWs. Since we do not know the polarization of the GWs that we are looking for, we use an approximation of the ideal filter function. We choose the unpolarized filter function  $\widetilde{Q}_{IJ}(t;f,\hat{\Omega}) = \frac{1}{\epsilon_{IJ}(t;\hat{\Omega})}e^{2\pi i f \hat{\Omega} \cdot \Delta \vec{x}_{IJ} / c} $\citep{ThKa2011}. The second term in this equation corrects for a difference in phase due to different travel distances from the source to the two detectors. For this discussion, we will set it to 1 for simplicity.

If the GWs emitted from magnetars are due to a pure quadrupole, we expect them to be polarized elliptically according to standard linearized gravity, with the polarization depending on the orientation of the source's time changing quadrupole moment with respect to our detectors. 

\begin{figure}[ht]
\includegraphics[width=0.48\textwidth]{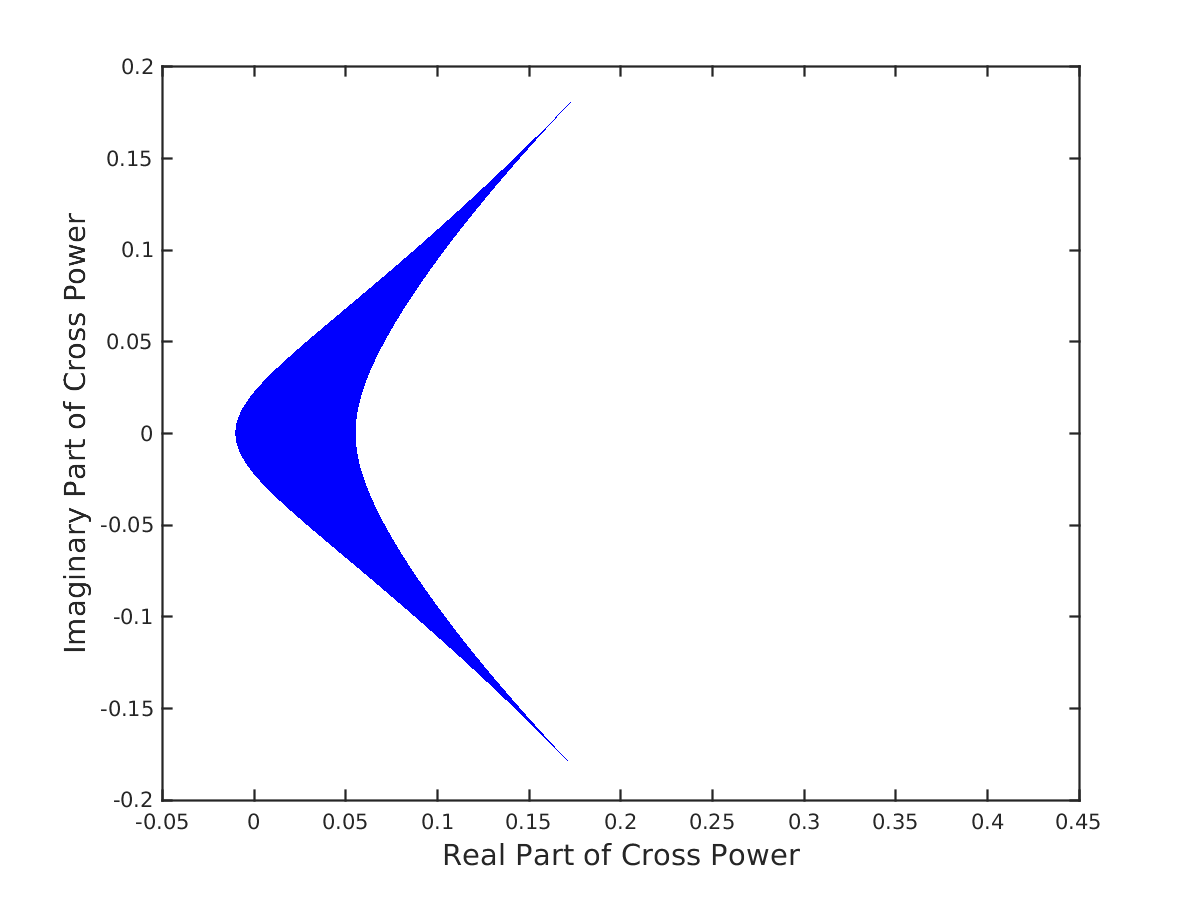}
\includegraphics[width=0.48\textwidth]{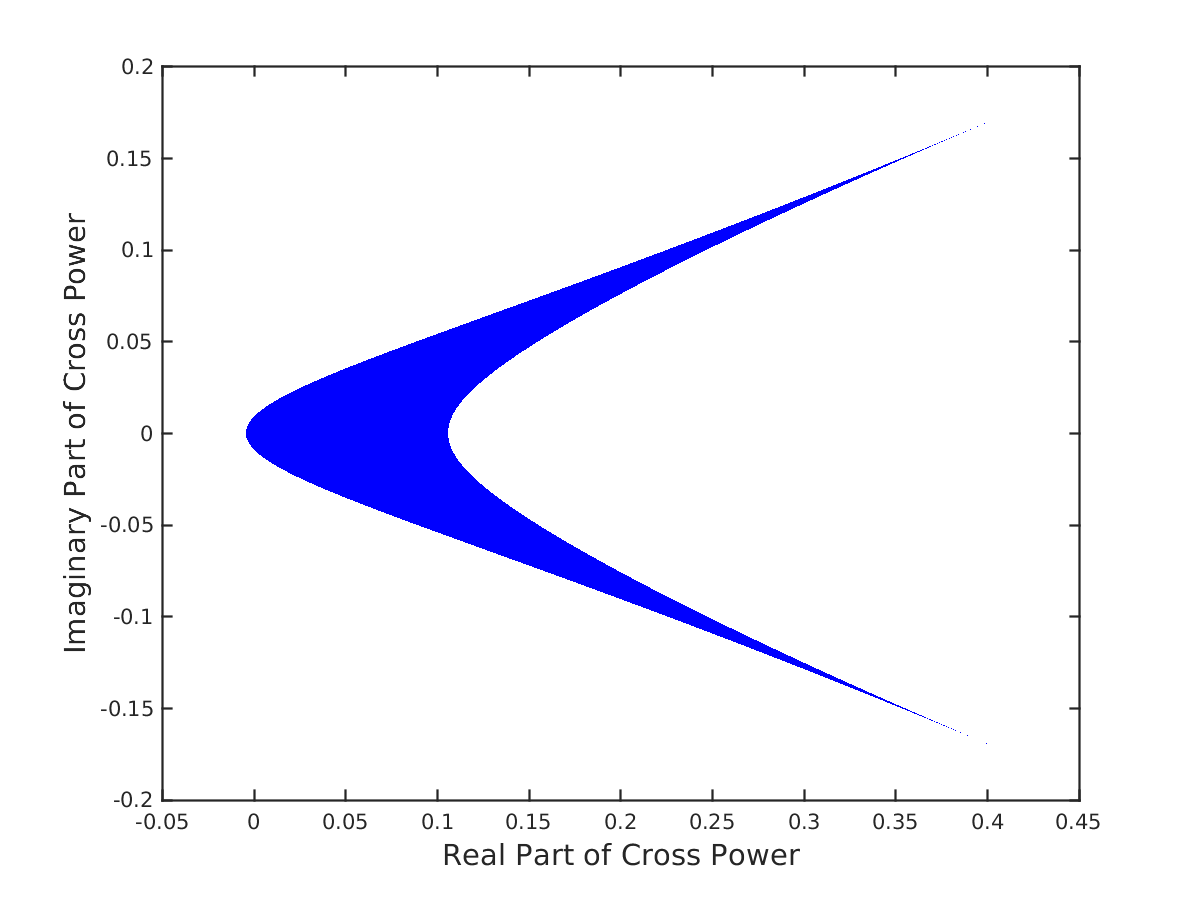}
\caption{\label{boomerangs}Parametric plots of the complex valued cross power due to elliptically polarized signals of varying polarizations from two different sky locations (left: SGR 1806-20 during the event on April 29, right: SGR 1806-20 during the event on February 25). The polarization of incoming GWs is defined by two angles, $\iota$ and $\psi$. $\iota$ is the angle between the vector from Earth to the source and the source's rotation vector, while $\psi$ indicates the orientation of the source's rotation vector when projected into a plane perpendicular to the propagation vector. The ends of the boomerang are at $\iota=0, \pi$; changing $\psi$ changes the real part only. For ideal sky positions, the boomerang collapses to the real axis and reaches about 0.95. It does not reach 1 because the detectors are not aligned.}
\end{figure}

Using Equation A48 from \cite{ThKa2011}, we can see how the cross power of an individual pixel will change with changing polarization. In Fig.~\ref{boomerangs}, we plot this parametrically in the complex plane for two of the bursts analyzed. The main effect of the filter function is to pick a phase (the cross power is multiplied by the filter function, then the real part is taken). The unpolarized filter function sets that phase to zero, so the imaginary part of the cross power is ignored, which limits the sensitivity to nearly all elliptically polarized waveforms. Because the time-frequency map is normalized by the average value of the power spectral density nearby in time, the magnitude of the filter function only matters if it changes over the course of a time-frequency map.

In addition, for some polarizations, the cross power is negative, which will result in pixels in the SNR map being negative. For this reason these signals, no matter how large in amplitude, would be missed by this search. However, these signals produce little cross power (as evidenced by how close they are to zero) such that the likelihood of detecting them even with the ideal filter function is low. It is possible to recover these by taking the absolute value of each cluster, but doing this is roughly equivalent to running the search twice with opposite phase filter functions. Thus, it will double the background.

\section{Supplementary Data} \label{swiftplots}

\begin{figure}
\centering
\includegraphics[width=0.5\textwidth, angle=90]{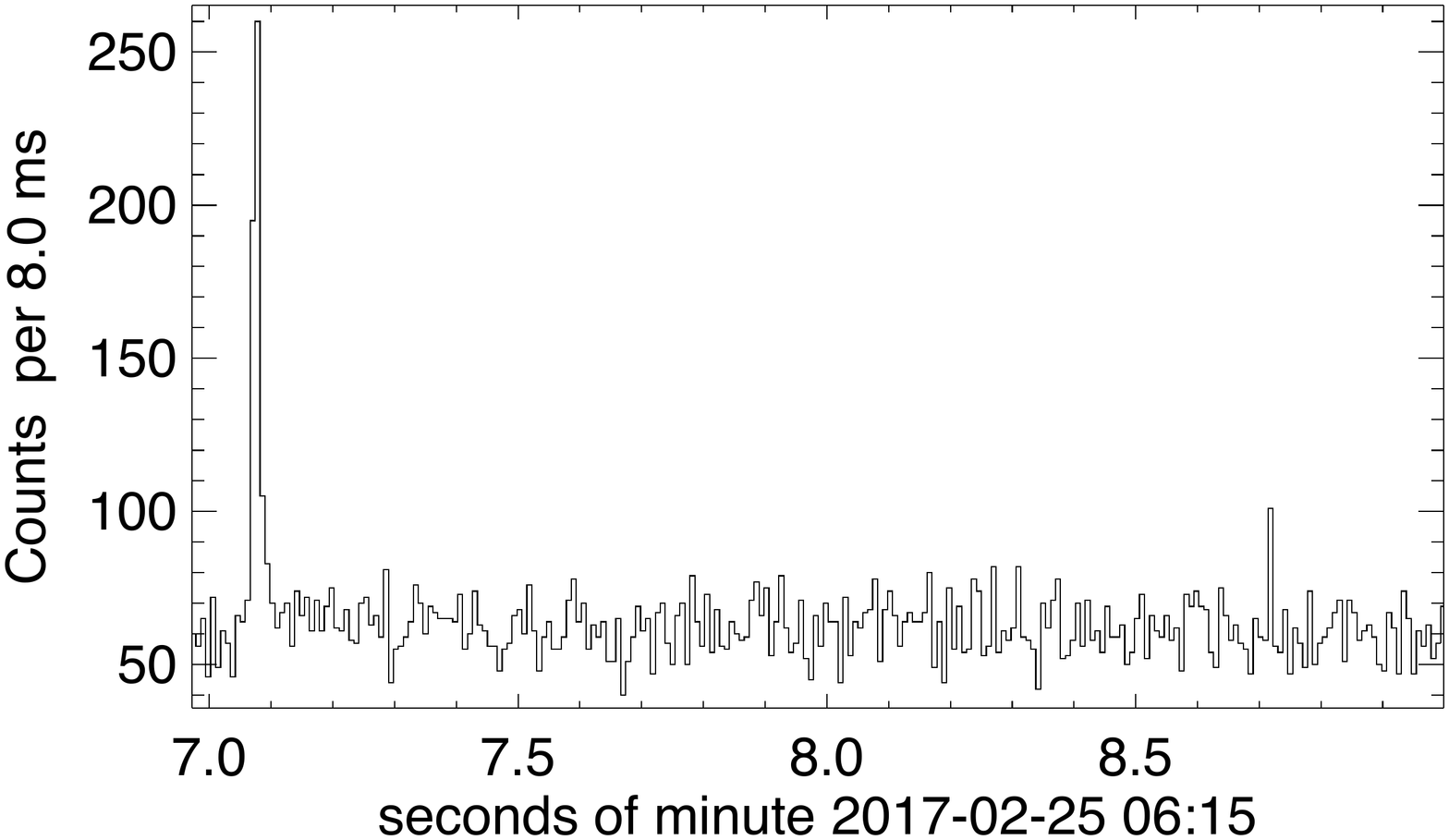}
\includegraphics[width=0.5\textwidth, angle=90]{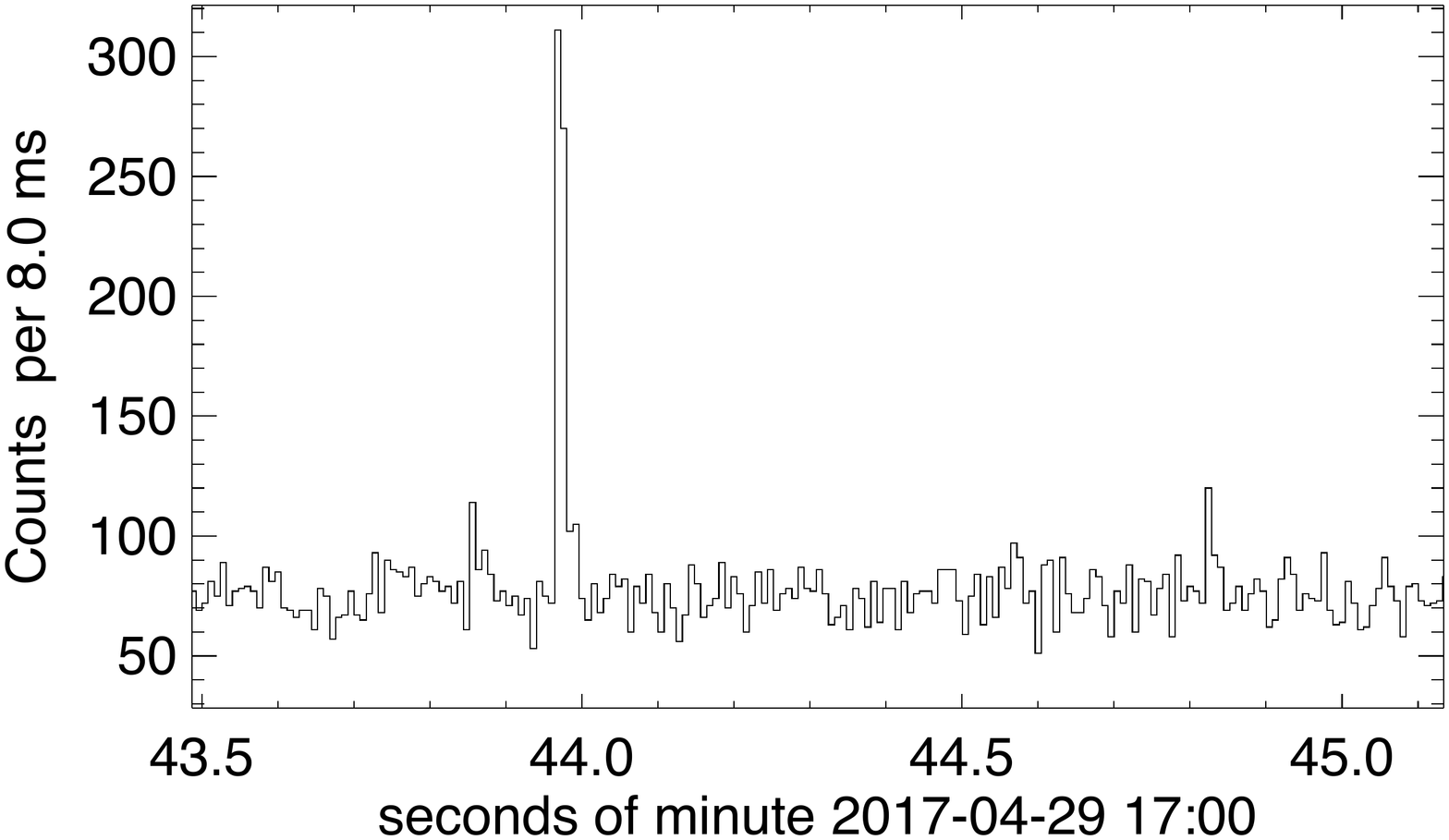}
\caption{\label{extra_swift_data}SWIFT BAT's data for the February 25 and April 29 bursts from SGR 1806-20. Images courtesy of David M. Palmer.}
\end{figure}

\iftoggle{endauthorlist}{
  %
  %
  \let\author\myauthor
  \let\affiliation\myaffiliation
  \let\maketitle\mymaketitle
  \title{Authors}
  \input{LSC-Virgo-Authors-Aug-2018-aas}
  \newpage
  \maketitle
}

\end{document}

%% file: LSC-Virgo-Authors-Aug-2018-aas.tex
\author{B.~P.~Abbott}
\affiliation{LIGO, California Institute of Technology, Pasadena, CA 91125, USA}
\author{R.~Abbott}
\affiliation{LIGO, California Institute of Technology, Pasadena, CA 91125, USA}
\author{T.~D.~Abbott}
\affiliation{Louisiana State University, Baton Rouge, LA 70803, USA}
\author{S.~Abraham}
\affiliation{Inter-University Centre for Astronomy and Astrophysics, Pune 411007, India}
\author{F.~Acernese}
\affiliation{Universit\`a di Salerno, Fisciano, I-84084 Salerno, Italy}
\affiliation{INFN, Sezione di Napoli, Complesso Universitario di Monte S.Angelo, I-80126 Napoli, Italy}
\author{K.~Ackley}
\affiliation{OzGrav, School of Physics \& Astronomy, Monash University, Clayton 3800, Victoria, Australia}
\author{C.~Adams}
\affiliation{LIGO Livingston Observatory, Livingston, LA 70754, USA}
\author{R.~X.~Adhikari}
\affiliation{LIGO, California Institute of Technology, Pasadena, CA 91125, USA}
\author{V.~B.~Adya}
\affiliation{Max Planck Institute for Gravitational Physics (Albert Einstein Institute), D-30167 Hannover, Germany}
\affiliation{Leibniz Universit\"at Hannover, D-30167 Hannover, Germany}
\author{C.~Affeldt}
\affiliation{Max Planck Institute for Gravitational Physics (Albert Einstein Institute), D-30167 Hannover, Germany}
\affiliation{Leibniz Universit\"at Hannover, D-30167 Hannover, Germany}
% ...
